# A3D-MoE: Acceleration of Large Language Models with Mixture of Experts via 3D Heterogeneous Integration

Wei-Hsing Huang, Janak Sharda, Cheng-Jhih Shih, Yuyao Kong, Faaiq Waqar, Pin-Jun Chen, Yingyan (Celine) Lin and Shimeng Yu, *Fellow, IEEE*

*Abstract*—Conventional large language models (LLMs) are equipped with dozens of GB to TB of model parameters, making inference highly energy-intensive and costly as all the weights need to be loaded to onboard processing elements during computation. Recently, the Mixture-of-Experts (MoE) architecture has emerged as an efficient alternative, promising more efficient inference with less activated weights per token. Nevertheless, fine-grained MoE-based LLMs face several challenges: 1) Variable workloads during runtime create arbitrary GEMV-GEMM ratios that reduce hardware utilization, 2) Traditional MoE-based scheduling for LLM serving cannot fuse attention operations with MoE operations, leading to increased latency and decreased hardware utilization, and 3) Despite being more efficient than conventional LLMs, loading experts from DRAM still consumes significant energy and requires substantial DRAM bandwidth. Addressing these challenges, we propose: 1) A3D-MoE, a 3D Heterogeneous Integration system that employs state-of-the-art vertical integration technology to significantly enhance memory bandwidth while reducing Network-on-Chip (NoC) overhead and energy consumption. 2) A 3D-Adaptive GEMV-GEMM-ratio systolic array with V-Cache efficient data reuse and a novel unified 3D dataflow to solve the problem of reduced hardware utilization caused by arbitrary GEMV-GEMM ratios from different workloads, 3) A Hardware resource-aware operation fusion scheduler that fuses attention operations with MoE operations to enhance hardware performance, and 4) MoE Score-Aware HBM access reduction with even-odd expert placement that reduces DRAM access and bandwidth requirements. Our evaluation results indicate that A3D-MoE delivers significant performance enhancements, reducing latency by a factor of 1.8× to 2× and energy consumption by 2× to 4×, while improving throughput by 1.44× to 1.8× compared to the state-of-the-art.

## I. INTRODUCTION

Decoder-only transformer Large Language Models (LLMs) such as GPT and Llama have demonstrated remarkable capabilities across diverse applications [1]. Over the past 5 years, the model parametric footprint has been increased by 3 orders of magnitude towards the TB level to enhance model quality. However, this brute-force scaling approach faces demanding computational challenges and memory bandwidth requirements. To address the aforementioned challenges, the Mixture-of-Experts (MoE) architecture, as demonstrated by Qwen and DeepSeek [2], has emerged as an efficient alternative, allowing models to scale parameters without proportionally increasing computational overhead during inference. In contrast to previous coarse-grained MoE structures, fine-grained MoE structures demonstrate superior performance characteristics and have emerged as the predominant architecture in the field [6]. This work focuses on accelerating fine-grained MoE architectural implementations.

MoE layers replace conventional feed-forward networks with multiple specialized expert networks and a gating mechanism for expert selections. Despite reduced computational demands, MoE-based LLMs still present significant memory bandwidth challenges for edge devices. Previous approaches include: 1. Utilize analog compute-in-memory (ACIM) [7] to reduce bandwidth requirements, but compromise accuracy. 2. Utilize Processing-in-memory (PIM) and bring computation closer to memory, but face scalability issues due to the performance limitations of DRAM peripheral transistors relative to their logic technology counterparts, resulting in higher latency, energy, and area compared to advanced nodes, as illustrated in the architecture presented in Fig. 1(a), Type-1 [8,9]. 3. Relocate Single Instruction, Multiple Data core (SIMD) units into the HBM logic die (duplex [10]), connected to a GEMM logic die via 2.5D interposer. While increasing bandwidth, this requires energy-intensive Serializer-Deserializer (SerDes) interfaces and forces GEMM units to retrieve data through extensive NoC, causing substantial energy consumption, as demonstrated in the architectural framework illustrated in Fig. 1(a), Type-2. On the hardware implementation front, 2.5D/3D heterogeneous integration has witnessed substantial progress through advancements in hybrid bonding and through-silicon-vias (TSV) that scale I/O pitch to sub-5µm [11], and industry products such as high-bandwidth-memory (HBM, 3D stacked DRAM) and V-Cache (3D stacked SRAM [12]) have reaped the benefits of vertical die stacking.

This work was supported in part by the PRISM, one of the SRC/DARPA JUMP 2.0 centers, and by the Department of Health and Human Services Advanced Research Projects Agency for Health (ARPA-H) under Agreement Number 140D042490003. The views and conclusions contained herein are those of the authors and should not be interpreted as necessarily representing the official policies or endorsements, either expressed or implied, of the Advanced Research Projects Agency for Health or the U.S. Government.

Wei-Hsing Huang, Janak Sharda, Yuyao Kong, Faaiq Waqar, Pin-Jun Chen, and Shimeng Yu are with the School of Electrical and Computer Engineering, Georgia Institute of Technology, Atlanta, GA 30332 USA (Corresponding author. E-mail: shimeng.yu@ece.gatech.edu).

Cheng-Jhih Shih, Yingyan (Celine) Lin are with the School of Computer Science, Georgia Institute of Technology, Atlanta, GA 30332 USA.

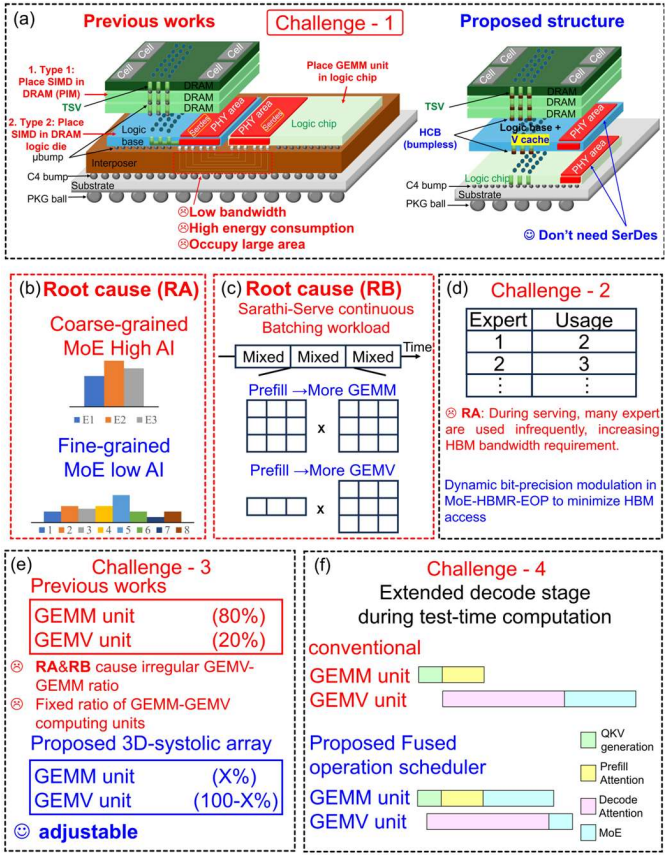

**Fig. 1.** Challenges hindering energy-efficient LLM inference.

significantly between low and high arithmetic intensity, leading to irregular GEMV-GEMM ratio operations. Note that even among GEMM operations, low arithmetic intensity GEMM operations cause extremely low utilization of GEMM computing units; this work refers to this phenomenon collectively as an irregular GEMV-GEMM ratio. Second, LLM inference serving is essential for delivering fast, scalable, and cost-effective inference capabilities with minimal latency while handling concurrent requests from multiple users. To enhance service quality for multiple requests, state-of-the-art approaches continue batching with Sarathi-Serve [14], which divides lengthy prefill sequences into multiple chunked-prefills to address the generation stall issues in ORCA [15] and implements stall-free scheduling. This results in most batching cases involving mixed prefill stage (more GEMM operations) and decode stage (more GEMV operations), as shown in Fig. 1(c). Consequently, previous work using fixed quantities of SIMD computing units and systolic array architectures results in inefficient hardware utilization under irregular GEMV-GEMM ratio conditions. This work proposes a 3D-Adaptive GEMV-GEMM-ratio systolic array (3D-systolic array) to address challenge 3. Furthermore, test-time computation has been empirically demonstrated to significantly enhance LLM performance metrics and has been implemented across numerous commercial conversational agents, including ChatGPT, Grok, Claude, and DeepSeek. Nevertheless, the utilization of test-time computation substantially increases the temporal requirements of the decoding phase. Traditionally, systems wait for all attention layers to complete before fetching experts, ensuring single DRAM access per expert. Loading experts prematurely based on partial results leads to eviction and reloading, increasing energy consumption and bandwidth usage. As shown in Fig. 1(f), our hardware resource-aware operation fusion scheduler (HR-OFS) enables concurrent QKV generation, attention and MoE operations, allowing GEMM units that complete prefill operations early to begin MoE operations without waiting for all decode stage operations to finish, significantly improving hardware utilization.

Advanced packaging enables ultra-high I/O vertical density (>100k/mm$^2$). As shown in Fig. 1 (a), our proposed structure vertically integrates a compute logic die with HBM using TSV technology, offering: 1. SerDes elimination by transitioning from coarse-grained 2.5D interposer technology to densely connected bumpless vertical connections, reducing DRAM access energy. 2. Direct data transmission via TSVs, minimizing NoC energy consumption. 3. 3D V-Cache architecture maximizing data reuse and minimizing communication overhead to the computing logic die. Our implementation also uses bumpless HBM technology with higher thermal conductivity (Cu) thanks to the adoption of hybrid bonding [13], therefore improving thermal dissipation and overall system performance.

As shown in Fig. 1(d), challenge-2 is caused by Root cause A, where fine-grained MoE structures contain significantly more experts than coarse-grained MoE structures. This results in selected experts being distributed across a wider range during batching operations, causing many experts to be utilized by only a small number of tokens, which leads to low arithmetic intensity (AI). The proposed MoE Score-Aware HBM access reduction with even-odd expert placement (MoE-HBMR-EOP) reduces overall HBM access counts by dynamically regulating whether to access full or half precision experts from HBM. As shown in Fig. 1(e), challenge-3 results from Root cause A and B. First, the fine-grained MoE structure contains a larger number of experts, which means that when batch size or input token count changes, MoE layer computations fluctuate

In this work, we propose A3D-MoE, a novel approach to handling the irregularity in fine-grained MoE model serving that specifically addresses the dynamic GEMV-GEMM ratio characteristics of these models and irregular memory bandwidth requirement. Our system adaptively responds to changing workload patterns at fine granularity, making real-time decisions about resource allocation based on current batch composition, expert activation patterns, and sequence lengths. We take advantages of 3D heterogeneous integration to design a 3D energy efficient hardware platform for fine-grained MoE structure for long decode stage scenario.

The primary contributions of this work are listed as follows:
- To address irregular GEMV-GEMM ratio's operations during LLM MoE serving, we propose 3D-Adaptive GEMV-GEMM-ratio systolic array with a novel unified 3D dataflow. This approach resolves computational inefficiencies caused by the irregularity while simultaneously reducing both latency and energy consumption caused by data communication overhead and processing bubbles in conventional 2D dataflow implementations. In addition, we employ V-Cache-like architecture that facilitates the decomposition of low arithmetic intensity GEMM operations into GEMV operations, while enabling efficient data reuse within the 3D hardware structure.
- We propose the hardware resource-aware operation fusion scheduler, this scheduler enables the fusion of attention operations with the MoE stage during LLM inference serving. When GEMM computing units complete the attention operations in the prefill stage earlier than decode stage, A3D-MoE can begin executing MoE operations without waiting for the complete computation of attention operations in the decode stage, thereby significantly enhancing hardware utilization.
- During the MoE operation, loading experts from HBM consumes substantial energy and places significant demands on HBM bandwidth. To address these challenges, we leverage the characteristics of the MoE gating network and propose MoE-HBMR-EOP, which reduces HBM accesses, thereby decreasing both HBM access energy and bandwidth requirements.
- Finally, we conduct extensive experiments on modern fine-grained MoE models at two different scales (7B, 15B), using multiple datasets and LLM serving configurations to comprehensively evaluate the A3D-MoE architecture. Our evaluation results demonstrate significant improvements in latency and energy consumption.

| Model | Attn. Type | Params Act./Total | Experts (per layer) Act./Total |
|---|---|---|---|
| OLMoE-1B-7B [3] | MHA | 1B/7B | 8/64 |
| DeepSeek-V2-Lite [4] | MLA | 2.4B/15.7B | 1(shared) + 6/64 |
| Qwen-1.5-MoE-A2.7B [5] | MHA | 2.7B/14.3B | 1(shared) + 2/60 |

Table I. Summary of fine-grained MoE models to be profiled.

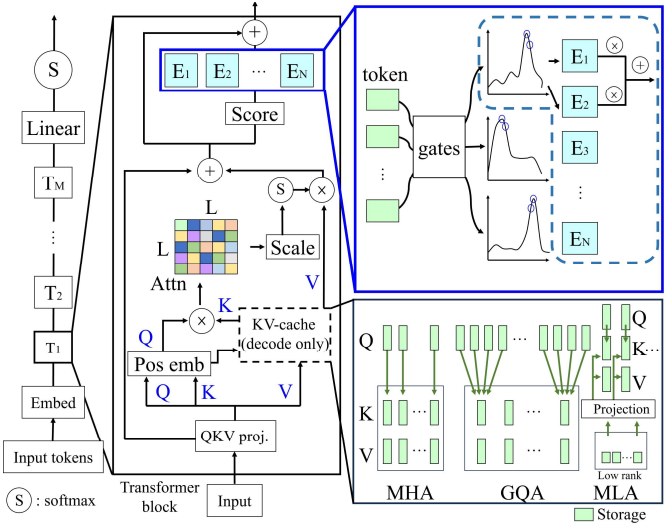

**Fig. 2.** Structure of the Mixture-of-Experts (MoE) LLM architecture.

## II. BACKGROUND

### A. *Decoder-Only Transformer LLM Architecture, Attention Mechanisms and Mixture-of-Experts*

1) **Decoder-only transformer LLM**
architectures consist of many cascaded transformer blocks, each combining a self-attention layer and a feed-forward network (FFN). Fig. 2 illustrates the MHA and FFN. A Multi-Head Attention (MHA) layer extends the standard self-attention mechanism by applying attention in multiple subspaces in parallel. Given a sequence of L input tokens $X \in R^{L \times D}$, MHA projects the inputs into multiple sets of queries, keys, and values:

$$Q_i = XW_i^Q, K_i = XW_i^K, V_i = XW_i^V \qquad (1)$$

where $i=[1..h]$ for each head *i*. Each head computes scaled dot-product attention independently:

$$S_i = softmax(Q_i K_i^T/\sqrt{d}), O_i = S_i V_i \qquad (2)$$

The outputs from all heads are then concatenated and projected through an output matrix:

$$MHA(X) = Concat(O_1, \ldots O_h) W^O \qquad (3)$$

Here, $W_i^Q, W_i^K, W_i^V \in R^{D \times d}$ are learned projection matrices, and $W_O \in R^{hd \times D}$ and $D = hd$. This architecture allows the model to jointly attend to information from different representation subspaces at different positions.

After the attention mechanism, FFN applies an up projection, a gating mechanism, and a down projection. Specifically, the input is projected to a higher-dimensional space, modulated by a gate (usually using a non-linear activation function like GELU or SwiGLU), and then projected back to the original hidden size. That is:

$$FFN(X) = \left(ACT(XW^G) \odot (XW^U)\right)W^D \qquad (3)$$

Here, $W^G, W^U \in R^{D \times D_{FFN}}$, $W^D \in R^{D_{FFN} \times D}$ and $D_{FFN}$ represents an intermediate dimension in the FFN.

2) **Advanced Attention Mechanisms**
MHA improves representational power but incurs high memory and bandwidth cost, since at inference time all key and value vectors for all heads must be stored (as KV caches) and loaded at each generation step. This has motivated several efficient attention variants to reduce overhead while preserving accuracy. **Grouped-Query Attention (GQA)** reduces memory and compute by allowing multiple query heads to share the same keys and values. Specifically, $h$ query heads are grouped into $g < h$ key-value sets. Each group shares $K_i, V_i$, while keeping separate $Q_i$. This lowers storage and bandwidth requirements during inference without significantly impacting model quality. **Multi-Head Latent Attention (MLA)** introduces a high-efficiency, low-rank key-value joint compression technique to eliminate the inference-time key-value cache bottleneck, thereby enabling efficient inference. This design improves efficiency and scalability, especially in long-context settings. We summarize the attention mechanisms in Fig. 2.

3) **Mixture-of-Experts (MoE)**
MoE replaces the standard FFN block with multiple expert sub-networks with each identical to a FFN block. Fig. 2 shows the whole MoE process. For each input token

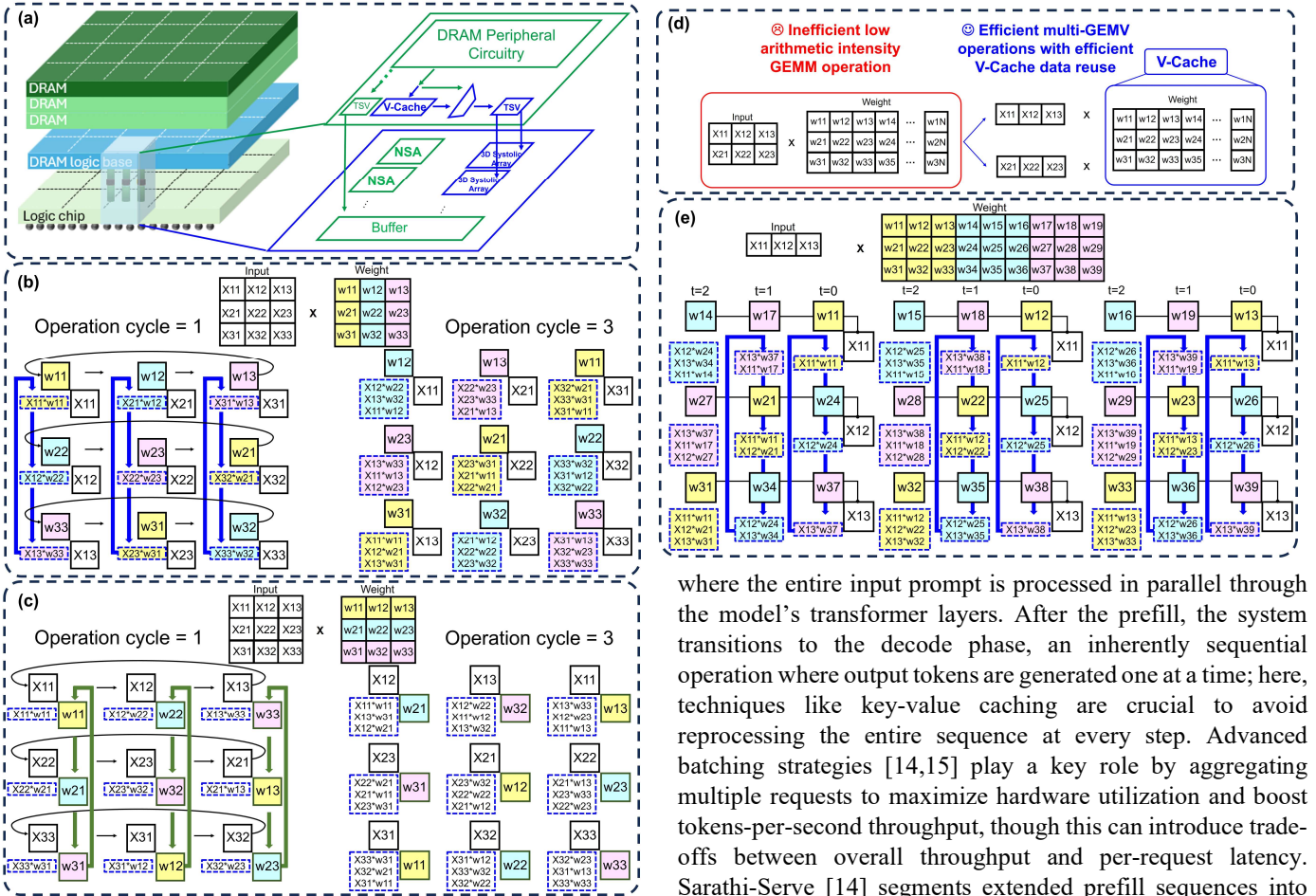

**Fig. 3.** The proposed hardware architecture and data flow for 3D-Adaptive GEMV-GEMM-ratio systolic array.

$x \in X$, a learned gating network computes routing weights:

$$G(x) = softmax(xW^g) \quad (4)$$

which selectively activates a sparse subset $S(x)$ of experts $\{E_1, ... E_N\}$. Each expert processes the input via its own projection matrices:

$$E_j(x) = \left(ACT(xW_j^G) \odot (xW_j^U)\right)W_j^D \quad (5)$$

and the final output is aggregated as a weighted sum:

$$MoE(x) = \Sigma_{j \in S(x)} G(x)_j E_j(x) \quad (6)$$

This approach boosts model capacity while only activating a few experts per token, keeping inference efficient.

Table I summarizes current popular MoE models selected for demonstration in this study, including their attention type, number of activated and total parameters, and number of experts. From this, we can see that many recent works (e.g. Qwen and DeepSeek) tend to adopt more fine-grained expert configurations.

### B. LLM Serving

LLM serving is the process of deploying large language models for efficient, scalable inference, and it requires a careful balance of computational throughput, memory management, and latency. The inference process begins with a prefill phase, where the entire input prompt is processed in parallel through the model's transformer layers. After the prefill, the system transitions to the decode phase, an inherently sequential operation where output tokens are generated one at a time; here, techniques like key-value caching are crucial to avoid reprocessing the entire sequence at every step. Advanced batching strategies [14,15] play a key role by aggregating multiple requests to maximize hardware utilization and boost tokens-per-second throughput, though this can introduce trade-offs between overall throughput and per-request latency. Sarathi-Serve [14] segments extended prefill sequences into multiple chunked-prefills to mitigate the generation stall issues observed in ORCA [15] and incorporated stall-free scheduling mechanisms. Consequently, Sarathi-Serve can achieve better throughput and low tail-latency compared to ORCA. Therefore, this work adopts Sarathi-Serve. However, Sarathi-Serve's approach of dividing lengthy prefill sequences into multiple chunked-prefills introduces new challenges. As discussed in the introduction, Sarathi-Serve causes hardware to operate predominantly in the mixed decode and prefill stage. The system's capability to effectively support this characteristic becomes essential for performance improvement.

### C. Related Works Accelerating LLM Inference

Several prior works have aimed to address the challenged posed by irregular GEMM-GEMV ratio computation pattern under limited memory bandwidth. The GEMM operation is characterized by high reuse of weights, inputs, and outputs, enabling low memory bandwidth requirements while achieving high FLOPS. In contrast, GEMV operations demand significantly higher memory bandwidth due to limited data reuse. To alleviate this issue, prior work [16] proposes analog compute-in-memory (ACIM) to directly reduce memory bandwidth requirements; however, this approach compromises LLM inference accuracy. Other works [8,9] leverage processing-in-memory (PIM) to bring computation closer to memory, thus maintaining accuracy. Nonetheless, this approach faces scalability issues, as the computing units in PIM must remain in the same technology node as DRAM (who's periphery devices are comparable to legacy logic technology nodes). To address the technology scaling limitation, [10] relocates the computation unit and placing it within the logic die of high-bandwidth memory (HBM). Nevertheless, this approach still utilizes computing units with a fixed GEMV-GEMM ratio, which remains insufficient for workloads characterized by irregular GEMV-GEMM ratio resulting from fine-grained MoE and Sarathi-Serve.

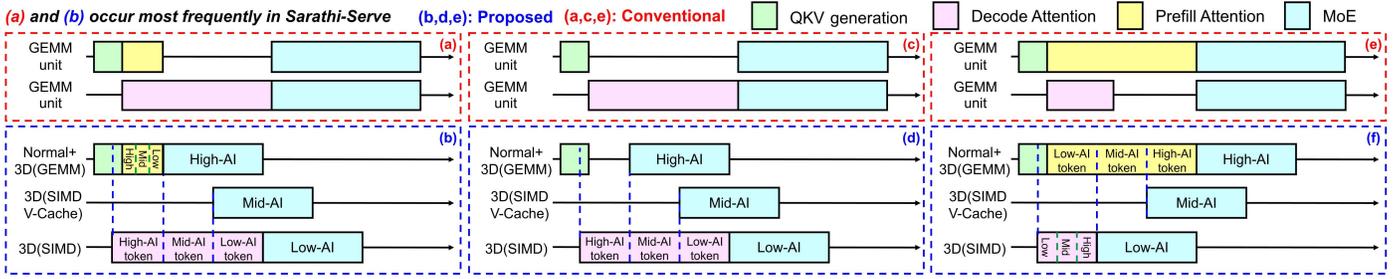

**Fig. 5.** The dataflow of the Hardware resource-aware operation fusion scheduler.

### III. 3   PROPOSED TOP-DOWN HW-SW CO-OPTIMIZATION

#### A. 3D-Adaptive GEMV-GEMM-ratio systolic array

As shown in Fig. 3, our proposed 3D architecture consists of multiple vertically stacked dies. At the bottom, a compute logic die, in the middle, an HBM base logic die, and at the top, multi-tier DRAM dies forming the HBM. Fine-Grained HBM [17] is adopted in this work to facilitate efficient routing, consequently resulting in diminished energy consumption. The systolic array in the compute logic die is divided into two types: normal systolic array (NSA) and 3D-Adaptive GEMV-GEMM-ratio systolic array (3D-Systolic Array). The normal systolic array accesses Type-1 SRAM on the compute die, where data is transferred from the upper HBM through TSVs. Meanwhile, we place an appropriate amount of Type-2 SRAM (V-Cache) on the DRAM logic die, allowing data to flow directly from either V-Cache or HBM through TSVs into the PEs of the 3D-Systolic Array below. Combined with our proposed 3D-Systolic array dataflow, this architecture: 1) enables runtime adaptability to arbitrary GEMV-GEMM ratios, achieving high utilization, and 2) reduces routing area overhead and consequently decreases runtime energy consumption through advanced 3D integration. In contrast to our proposed 3D-Systolic array, previous work maintains fixed hardware resource allocation between systolic array and SIMD once the hardware configuration is established, resulting in significant performance fluctuations when the workload's GEMV-GEMM ratio changes. Traditional systolic arrays exhibit extended pipeline filling latency when executing GEMM operations. Furthermore, when performing GEMV operations, an $N \times N$ systolic array demonstrates significantly reduced hardware utilization, falling to merely $1/N$ of its capacity. The 3D-Systolic Array is proposed to address these limitations. For clarity, our explanation will illustrate using a GEMM with shape 3 and a GEMV with shape 3 as examples. Please note that this dataflow can be generalized to GEMM and GEMV operations of arbitrary dimensions.

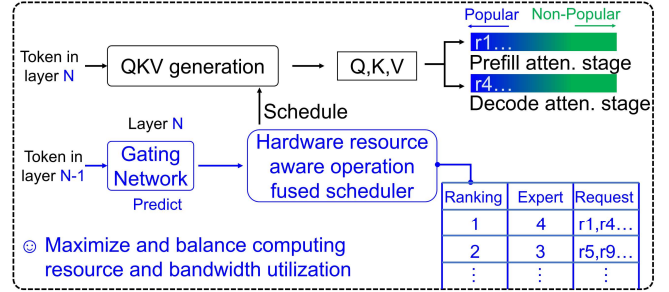

**Fig. 4.** The Hardware resource-aware operation fusion scheduler flow.

#### 1) GEMM-Weight stationary and input stationary dataflow (3D-GEMM)

Fig. 3(b) illustrates input stationary as an example; please note that the weight stationary dataflow follows the same principles. During execution, in the first cycle (loading cycle-1), all inputs enter the systolic array in parallel from the top die via TSVs. In contrast, traditional systolic arrays require N cycles to fully load inputs into the systolic array, resulting in extended pipeline filling latency. In the second cycle (loading cycle-2), all weights enter the systolic array in parallel from the top die via TSVs, whereas traditional systolic arrays need N cycles for weight transmission. Traditional systolic arrays implementing input stationary dataflow utilize temporal skewing of weights to ensure the proper accumulation of partial outputs during downward propagation. However, this approach increases pipeline filling latency during weight transmission. This work proposes an interleaving data streaming flow to address this limitation. As shown in Fig. 3(b), our approach spatially pre-skews the weight positions and transmits them directly through TSVs to the systolic array in a single cycle. Due to this spatial interleaving of weights, partial outputs can propagate downward each cycle and accumulate with the correct partial outputs, similar to traditional systolic arrays. During each cycle, weights propagate rightward while partial outputs propagate downward, as in conventional systolic arrays. Unlike traditional implementations, our design requires the output buffer of row N to connect with the buffer of row 1, and the weight buffer of column N to connect with column 1. The final outputs are transmitted upward via TSVs in a single cycle, adjusted to normal output order format through a de-interleaving circuit, and then stored back to V-Cache or HBM.

#### 2) GEMM-Output stationary dataflow(3D-GEMM)

As illustrated in Fig. 3(c), during operation, all inputs enter the systolic array in parallel via TSVs from the top die during the first cycle (cycle-1). In the second cycle (cycle-2), all weights are transferred in parallel from the top die to the systolic array through TSVs. As described in Section 3.1.1, to ensure that partial outputs remain stationary within PEs and accumulate with the correct partial outputs, we perform spatial interleaving of weights prior to computation. During each cycle, inputs propagate rightward while weights propagate downward, with partial outputs remaining stationary within PEs for accumulation. The final outputs are transmitted upward via TSVs in a single cycle and stored back to either the V-Cache or HBM.

3) **GEMV(3D-GEMV and 3D-GEMV-Vcache)**

In LLM architecture, the attention layers and MoE layers within the decode stage involve GEMV operations. Previous work has allocated fixed quantities of SIMD units either in the logic die or implemented PIM to execute GEMV operations. However, when workload characteristics change, altering the GEMV-GEMM ratio, the utilization of these SIMD units experiences significant fluctuations. To address these challenges: 1. The proposed 3D-systolic array decomposes low arithmetic intensity GEMM operations into multiple GEMV operations and achieves efficient data reuse with the V-Cache in the 3D dimension, as shown in Fig. 3(d). 2. The proposed 3D-Systolic Array not only accelerates GEMM but also GEMV operations. As illustrated in Fig. 3(e), during GEMV execution, vector (input) parallelism is achieved along the x direction. Traditional SIMD approaches complete multiple input-weight multiplications and accumulate the results within a single cycle, resulting in extended critical paths that increase both latency and energy consumption. As mentioned in 3.1.1, our proposed data interleaving resolves this challenge. To ensure that partial outputs can propagate downward each cycle and accumulate with the correct partial outputs, we implement weight interleaving. After three cycles, all PEs obtain their final outputs. These final outputs are transmitted upward via TSVs in a single cycle, restored to the normal output order format through a de-interleaving circuit, and subsequently stored back to V-Cache or HBM.

B. *Hardware resource-aware operation fusion scheduler*

Previous work requires completion of all prefill and decode stage attention operations before MoE layer execution. After gating network scoring, top-$k$ experts are selected to begin MoE layer processing. Limited on-chip SRAM necessitates expert eviction when loading new ones. Conventional scheduling thus delays MoE layers until all attention layers finish, maximizing expert reuse (Fig. 5(a,c,e)). Test-time compute enhances LLM performance but increases decode stage proportion, causing GEMM units to stall after prefill computation, reducing utilization. Additionally, advanced continuous batching techniques such as Sarathi-Serve [14] cause systems to predominantly enter extended decode-intensive mixed prefill and decode stages, as demonstrated in Fig. 5(a). Prior works [18,19] demonstrate that expert usage can be predicted within

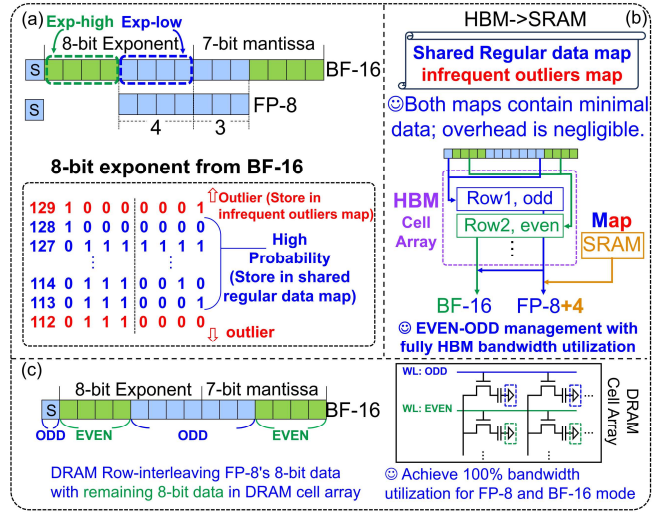

**Fig. 6.** The proposed MoE-HBMR-EOP framework.

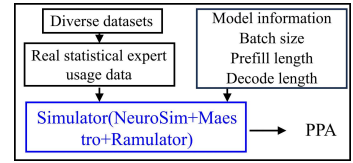

**Fig. 7.** The proposed simulation workflow

**Table III.** (a) Distribution of exponents. (b) Accuracy analysis of MoE-HBMR-EOP.

(a)
| Model | 4-bit Exponent range | Percentage |
|---|---|---|
| OLMoE | 107~122 | 99.94% |
| DeepSeek | 112-127 | 99.87% |
| Qwen | 106~121 | 99.94% |

(b) BA.: Baseline accuracy
| Model | BA. | Acc. With MoE-HBMR-EOP |
|---|---|---|
| OLMoE | 54.1 | 54.1 |
| DeepSeek | 58.3 | 58.2 |
| Qwen | 61.0 | 60.9 |

| | A3D-MoE-1(Setting-1) | A3D-MoE-2 (Setting-2) |
|---|---|---|
| Normal systolic Array | 32x32x384 | 32x32x768 |
| 3D Systolic Array | 16x16x512 | 16x16x1024 |
| # HBM | 1 | 2 |
| HBM Capacity | 36 GB | 72 GB |
| Bandwidth per HBM | 9600 GB/s (enabled by fine pitch TSV / bumpless HBM) | |
| Type-1 SRAM Capacity | 16 MB | 32 MB |
| Type-2 SRAM Capacity(V-cache) | 16 MB | 32 MB |
| Area (mm²) (Same as HBM logic die) | 121 | 242 |
| Frequency | 1 GHz | |
| Technology | 7 nm | |

**Table II.** The hardware parameters of A3D-MoE.

>90% accuracy by feeding layer $i$-1 tokens into layer $i$'s gating network without retraining. This addresses scenarios where HBM cannot store all experts, but doesn't explore cases where GPU HBM is sufficient. Our paper tackles scenarios with adequate on-chip HBM, eliminating pre-fetch needs. We propose a hardware resource-aware operation fusion scheduler (HR-OFS) to maximize bandwidth and computing utilization during inference serving.

HR-OFS (Fig. 4) ranks experts by token usage frequency as high or low arithmetic intensity (AI) experts and identifies bottlenecks through prefill/decode analysis. When the decode stage dominates (memory-bound), we prioritize high AI-MoE operations; conversely, when prefill dominates (compute-bound), we prioritize low AI-MoE operations. For decode-dominated processes (Fig. 5(b)), HR-OFS prioritizes QKV generation for high AI-tokens, enabling immediate attention operations while maximizing hardware utilization. QKV generation and prefill attention being compute-bound, we complete QKV generation before prefill attention. When prefill attention is completed early and the computation of high AI-tokens in the decode stage is finished, it enables subsequent concurrent execution of compute-bound High AI-MoE operations and memory-bound decode attention, thereby optimizing overall utilization. High-AI GEMM operation (High-AI MoE) uses 3D-systolic arrays in GEMM mode; low-AI GEMM operation (Mid-AI MoE) decomposes into multi-GEMV utilizing the V-Cache for data reuse. Low-AI MoE operations use GEMV mode after decode completion. As illustrated in Fig. 5(d), the computational flow of the decode-only stage exhibits substantial similarity to that of the decode-dominant mixed prefill and decode stages, following analogous processing patterns but with the notable absence of prefill computation operations. For prefill dominance (Fig. 5(f)), prioritizing low AI-tokens during QKV generation allows low AI-MoE operations to be executed immediately after both the decode phase is complete and the computation of low-AI tokens in the prefill stage is finished, allowing concurrent operation of compute-bound prefill attention and memory-bound Low AI-MoE operations. The 3D-systolic array (Fig. 5(b,d,f)) is partitioned into 3D-GEMM, 3D-SIMD, and 3D-SIMD-V-Cache with run-time reconfigurable ratios, automatically switching modes when components are idle. Due to low prediction accuracy in layers 1-3 [18,19], HR-OFS implementation begins from layer 4.

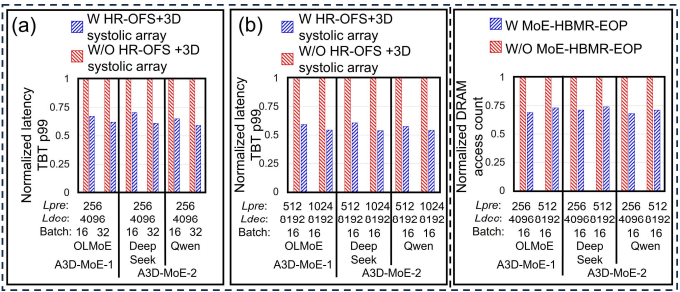

**Fig. 8** Comparison of Normalized latency (TBT P99) between proposed HR-OFS and conventional method.

**Fig. 9** Comparison of Normalized DRAM access count.

### C. MoE Score-Aware HBM access reduction with even-odd expert placement

When computing MoE operations, non-popular expert MoE operations exhibit limited data reuse, resulting in low arithmetic intensity and consequently consuming significant DRAM bandwidth. To address this challenge, this work proposes MoE Score-Aware HBM access reduction with even-odd expert placement (MoE-HBMR-EOP). During MoE operations, each MoE path's output is ultimately multiplied by its corresponding gating score and then aggregated. We observe that for a given token, among the selected top-$K$ experts, a small subset of experts typically dominates the overall score, while the remaining experts contribute minimally. Based on this characteristic, when accessing non-popular experts from DRAM with scores below a predefined threshold, MoE-HBMR-EOP retrieves only FP-8 representation instead of the original BF-16 precision, thereby reducing DRAM access energy. However, since the exponents in BF-16 and FP-8 formats are not aligned, conventional approaches would require reading the 8-bit BF-16 exponent and adjusting the bias before converting to the FP-8 exponent. This process would still necessitate reading 16 bits from HBM. Through profiling multiple MoE LLMs, we discovered that the exponent values in MoE layers have a notably narrow range, with most exponents representable in 4 bits. Consequently, we analyze the exponent distribution of each model layer offline to identify value ranges that maximally cover all exponent values for that layer. This information is stored in HBM and subsequently loaded into SRAM during runtime execution. As illustrated in Fig. 6(a,b), the shared Regular data map information in HBM indicates that when using FP-8 mode to read a 4-bit exponent (exp-low) for Regular data, if the exp-low is 4'b0000, then the exp-high will be 4'b1000; otherwise, if the exp-low is not 4'b0000, the exp-high will be 4'b0111. Most Regular data can share this information, rendering the overhead negligible. Additionally, HBM addresses and exp-high information for the small number of outlier data points are stored separately in an outlier map in HBM. Due to their limited quantity, this overhead is also negligible. Both maps are loaded directly into SRAM at runtime to ensure efficient execution of MoE-HBMR-EOP. This look-up process runs on the HBM logic die or compute logic die in advanced process nodes. Compared to the energy consumed by DRAM access, the energy overhead of this conversion process is minimal. The converted FP-8 values obtain the the correct and complete 8-bit exponent values and, after zero-padding the mantissa, are input to the BF-16 computing unit for computation. To maximize HBM bandwidth utilization, we split the BF-16 data format into two parts: the FP-8 portion is stored in the odd rows of the DRAM cell array, while the remaining BF-16 data is stored in the even rows. This placement ensures that whether accessing FP-8 or full BF-16 data, the HBM bandwidth can be fully utilized, as shown in Fig. 6(c).

## IV. EVALUATION RESULTS

### A. Methodology

For evaluation methodology, we conduct an ablation study of the proposed schemes in Sections 4.2 and 4.3 followed by a comprehensive comparison in Section 4.6 between A3D-MoE and two baseline implementations: a NeuPIM-based accelerator [8] and a Duplex-based accelerator [10]. During comparison, we maintain the SIMD flops allocated for HBM in the original NeuPIM paper and the SIMD flops allocated for HBM logic die in the original Duplex paper. Furthermore, since NeuPIM and Duplex employ 2.5D interposer connections to the logic chip, this incurs additional SerDes area overhead. However, for fair comparison purposes, we assume that the computational logic chip area utilized by NeuPIM and Duplex is larger than in our proposed architecture, in order to maintain an equivalent number of GEMM computation units, type-1 plus type-2 SRAM capacity, and identical DRAM bandwidth specifications. Detailed hardware parameters are enumerated in Table II. Based on model scale, we configure A3D-MoE in two variants: A3D-MoE-1, A3D-MoE-2, specifically designed to evaluate three fine-grained MoE-based LLMs. Additionally, when comparing Duplex and NeuPIM with A3D-MoE-2, identical hardware parameters to those of A3D-MoE-2 are maintained (setting-2) to ensure equitable evaluation. As illustrated in Fig. 7, we construct a comprehensive hardware system cycle-accurate simulator based on NeuroSim [20] and Maestro [21], integrated with Ramulator [22] to simulate overall HBM performance. Leveraging the proposed 3D-systolic array architecture, we implement hardware using Verilog and perform synthesis, automatic place and route (APR), and post-simulation utilizing a commercial 7 nm process design kit (PDK) to simulate overall hardware performance. Additionally, we employ Ansys tools to analyze the limitations of heat dissipation of the heterogeneously integrated hardware system, ensuring the proposed architecture remains unaffected by thermal constraints. We select three fine-grained expert LLMs to evaluate A3D-MoE performance: (1) MLMoE-1B-7B, (2) DeepSeek-V2-Lite (3) Qwen-1.5-MoE-A2.7B. To account for authentic data distribution variations across experts in fine-grained MoE implementations, we utilize diverse datasets including MMLU [23], MATH-500 [24] and Livecodebench [25] to statistically analyze expert utilization probability across MoE layers for each evaluated model. This approach prevents discrepancies between expert utilization statistics and real-world user scenarios that might arise from continuously inputting highly similar requests within sequences. These expert utilization statistics are subsequently fed into our implemented scheduler, which, based on configured prefill token length, decode stage length, and batch size parameters, dispatches requests to the hardware system cycle-accurate simulator according to Poisson distribution principles to evaluate comprehensive hardware performance. In the ablation studies presented in Sections 4.2 and 4.4 of the evaluation, to quantify the performance impact of individual proposed techniques, we maintain the data precision at the original precision of the selected models (BF-16). MoE-HBMR-EOP is subsequently incorporated into the system-level evaluation in Sections 4.3 and 4.6, where the gating network score dynamically determines whether to access BF-16 or FP-8 representations from HBM. In subsequent experiments, we employ $L_{pre}$ to denote the prefill token length and $L_{dec}$ to denote the decode length. During testing, all $L_{dec}$ values exceed $L_{pre}$ to simulate test-time compute conditions.

### B. Hardware resource aware operation fused scheduler and 3D-Adaptive GEMV-GEMM-ratio systolic array

As Hardware resource-aware operation fusion scheduler (HR-OFS) was co-designed with the 3D-systolic array, the full performance enhancement of HR-OFS depends on the 3D-systolic array's capability to dynamically transition between GEMM-GEMV operations. Therefore, this section analyzes the combined performance improvements of HR-OFS and the 3D-systolic array. To quantify the benefits of the proposed techniques, we compare HR-OFS + 3D-systolic array with conventional scheduling, which requires the completion of all attention operations before executing MoE layer computations. As this work focuses on scenarios where test-time computation induces prolonged decode stages, the 99[th] percentile token-between-token latency (TBT p99) most effectively reflects hardware performance stability [14]. Thus, we selected this metric for our ablation study. Fig. 8(a-b) demonstrates the effectiveness of our method. For three models with corresponding hardware settings, HR-OFS reduces latency by 1.42× to 1.86×. This reduction stems from addressing a critical bottleneck during test-time computation, where decode attention stages significantly expand, causing the system's GEMM units to enter a stall stage until decode attention completes, before MoE layer operations can execute—substantially increasing latency. HR-OFS, however, fuses attention operations with MoE operations, mitigating this latency penalty. Furthermore, as shown in Fig. 8, performance improvements from HR-OFS increase with larger batch sizes or expanded $L_{pre}$, as these conditions enable more tokens to operate on MoE layers, elevating expert reuse rates. When the prefill stage attention is completed, high arithmetic intensity MoE operations can commence shortly thereafter, as illustrated in Fig. 5(b).

### C. MoE Score-Aware HBM access reduction with even-odd expert placement

To evaluate MoE Score-Aware HBM access reduction with even-odd expert placement (MoE-HBMR-EOP), we first profiled the model and analyzed the value distribution of exponents in MoE experts across different models, including the probability of outlier occurrence. Table III(a) illustrates the coverage rate of 4-bit exponents across various models, demonstrating that most exponents fall within the 4-bit range, thus requiring minimal memory allocation for outlier map. For comparison purposes, we established the baseline by disabling MoE-HBMR-EOP. We first apply min-max normalization to rescale the raw gating scores into the [0,1] interval. After normalization to this range, our experiments revealed that utilizing FP-8 when gating scores are below 0.45 does not degrade accuracy, as shown in Table III(b). This preservation of accuracy is attributed not only to the selective use of FP-8 for lower-score elements but also to our implementation of LUT-based restoration for extreme outliers prior to computation, preventing accuracy degradation. Consequently, we established the threshold at 0.45. Fig. 9 demonstrates that our scheme reduces DRAM access counts by 1.35× to 1.44× across different test cases. Our approach not only reduces DRAM access counts but also lowers energy consumption during NoC data transmission, as shown in the energy analysis in Section 4.5, System-level comparison.

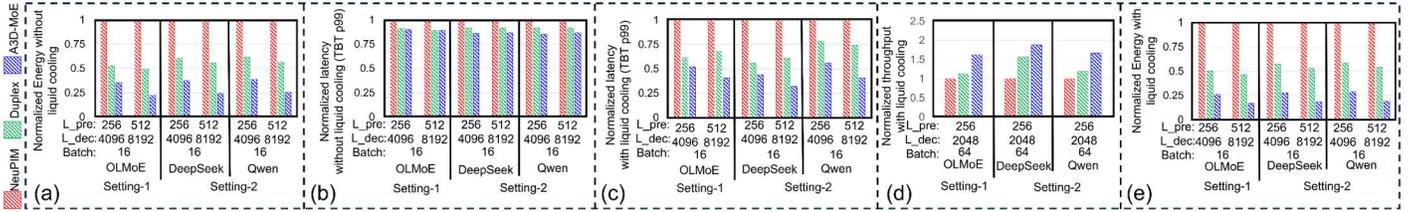

**Fig. 11.** Comparison of energy, latency (TBT), throughput between proposed A3D-MoE, duplex and NeuPIM with and without liquid cooling.

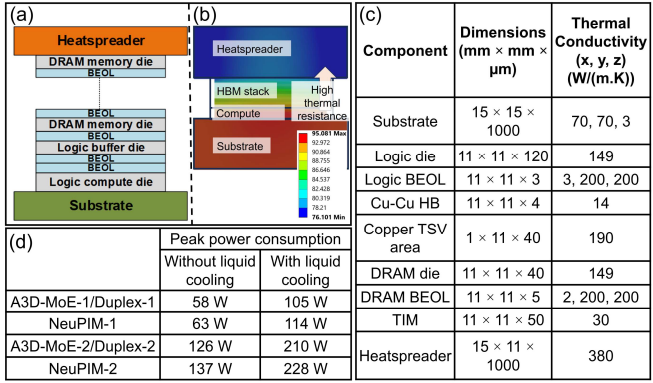

Fig. 10 (a) Materials stack used in thermal simulation (b) Thermal gradient (c) Component parameters and associated thermal conductivity (d) Permissible peak power consumption for different hardware, with and without liquid cooling.

### D. Area overhead estimation

Due to the prevalence of low arithmetic intensity GEMM operations in fine-grained MoE LLMs, using smaller systolic arrays improves flexibility and hardware utilization. We adopt a 16×16 3D-systolic array, which connects column/row 1 and N and achieves 1 GHz operation with minimal area cost thanks to its reduced dimensions. The 3D packaging also significantly cuts the routing area. Despite added overhead from the 3D-Systolic Array design, our analysis shows a ~10% area reduction in the compute logic die. This is due to direct vertical communication between the HBM and compute logic dies, reducing NoC area. For the HBM logic die, eliminating SerDes circuits frees area that we reuse for V-Cache and extra TSVs. Using advanced TSV technology, their combined area matches that of the removed SerDes, adding no extra area cost to the HBM logic die. In addition, the 3D-Systolic Array connects directly to the V-Cache, enabling efficient data reuse for low-AI GEMM. During high arithmetic intensity GEMM execution, the 3D-systolic array outperforms conventional systolic arrays due to significantly reduced pipeline filling time, as shown in Fig. 3. Furthermore, the reduction in NoC routing overhead contributes to lower energy consumption.

### E. Thermal evaluation

Next, we perform thermal simulations for the demonstrated co-packaged systems in Ansys Mechanical. The tight integration of logic and memory in a 3D stacked system leads to thermal coupling/management issues. Further, the high thermal resistance of the 12-heterogeneously integrated (HI) stack of the HBM restricts the heat dissipation from the bottom logic die to the top heat spreader. We consider two scenarios, with and without advanced liquid cooling, by varying the heat convection coefficient on the heatspreader. The peak operating power of the overall system is decided by making sure that the peak operating temperature for the DRAM dies is below 95 °C, as per the JEDEC specifications. Fig. 10(a) describes the schematic used for the thermal simulation, and Fig. 10(b) describes the obtained thermal map for a single 3D stack system. Fig. 10(c) describes the thermal parameters used for various components. Power consumption of the complete system is varied to obtain the peak temperature and a peak power consumption is obtained by limiting the peak temperature. For NeuPIM, we modify the power map by assigning power consumption to the compute on the DRAM logic to obtain the thermal map. For Duplex, the peak power limits are considered the same. For setting-2, we extend the system to a larger logic die connected to two HBM stacks and perform thermal evaluations. Fig. 10(d) describes the operating conditions for the two settings with and without liquid cooling. The obtained peak operating power consumption is used to evaluate the overall system-level performance.

### F. System-level comparison

As referenced in Section 4.1, for a fair comparison, we assume that Duplex has the same HBM bandwidth as our approach. Because Duplex involves computation integrated in the HBM logic die, the solution results in heat buildup in the HBM logic die and limits the peak performance. Conversely, NeuPIM incorporates SIMD units within DRAM, and due to the relatively less advanced logic units in the DRAM manufacturing process compared to logic fabrication, executing identical operations consumes greater energy, thus also causing thermal issues in the HBM during full-speed operation. Our experiments, therefore, compare scenarios with and without liquid cooling, where systems must reduce frequency without cooling to prevent thermal violations. Fig. 11(a) shows that A3D-MoE consistently achieves minimal energy consumption without liquid cooling across various conditions. This efficiency stems from our vertically integrated architecture, reducing routing and HBM access energy, while V-Cache enables efficient 3D data reuse for more effective execution of low arithmetic intensity GEMM operations, resulting in approximately 1.9× energy reduction compared to Duplex and 3.4× reduction versus NeuPIM. Fig. 11(b) demonstrates that without liquid cooling, performance throttling extends latency across all systems. As illustrated in Fig. 11(c), with liquid cooling, A3D-MoE's 99th percentile token-between-token latency decreases approximately 2× versus NeuPIM, attributable to the utilization of advanced TSV technology to enhance HBM bandwidth. During mixed decode and prefill stages, while NeuPIM struggles with irregular GEMM-GEMV operations, A3D-MoE reduces latency through HR-OFS and 3D-systolic arrays. Compared to Duplex, at minimal $L_{pre}$ values where MoE experts experience lower reuse rates, operations maintain arithmetic intensity marginally exceeding unity—precisely where Duplex architecture excels—yielding modest latency improvements. However, as $L_{pre}$ increases, expert reuse rates in mixed stages escalate, generating diverse irregular GEMM-GEMV operations that A3D-MoE efficiently executes via 3D-systolic arrays, processing high arithmetic intensity MoE operations after prefill stage completion, achieving up to 1.8× latency improvement. Fig. 11(d) presents a comparative analysis of throughput, exhibiting trends similar to those observed in Fig. 11(c). The efficiency of handling mixed decode and prefill stages is particularly crucial in Sarathi-Serve's chunked-prefills serving methodology. A3D-MoE demonstrates significant advantages in this phase, primarily benefiting from HR-OFS's capability to efficiently execute High AI-MoE and Mid AI-MoE operations following the prefill stage completion, as depicted in Fig. 5(b), while leveraging MoE-HBMR-EOP to reduce DRAM access latency. Consequently, A3D-MoE achieves throughput improvements of 1.6× to 1.8× compared to NeuPIM, and 1.2× to 1.44× relative to Duplex. Fig. 11(e) demonstrates A3D-MoE's excellent energy reduction performance, achieving an average 2× energy consumption decrease compared to Duplex and an average 4× reduction relative to NeuPIM. This substantial energy efficiency improvement is primarily attributed to diminished routing energy through 3D structural integration, elimination of 2.5D interposer architecture and associated SerDes circuits, complemented by MoE-HBMR-EOP's reduction in DRAM access frequency and HR-OFS with 3D-systolic array's enhancement of hardware utilization.

## V. CONCLUSION

This paper presents an innovative hardware-algorithm co-design framework for efficient LLM inference on resource-constrained devices. Our approach addresses challenges in state-of-the-art fine-grained Mixture-of-Experts (MoE) architectures and the complexities of mixed prefill and prolonged decode stages through three key innovations: 1. 3D-Adaptive GEMV-GEMM-ratio systolic array, which dynamically adapts to variations in GEMV-GEMM ratios through run-time mode switching; 2. Hardware resource-aware operation fusion scheduler (HR-OFS), which fuses attention and MoE operations to reduce latency and enhance overall hardware utilization; 3. MoE Score-Aware HBM access reduction with even-odd expert placement (MoE-HBMR-EOP), which reduces HBM access counts. Experimental results demonstrate that our proposed A3D-MoE achieves 1.8× to 2× latency reduction, 2× to 4× energy reduction, and 1.44× to 1.8× throughput improvements compared to the state-of-the-art.

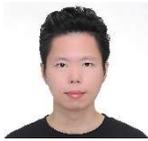
Wei-Hsing Huang received the B.S. degree in electrical engineering from the National Chung Cheng University, Chiayi, Taiwan, in 2017, and the M.S. degree in electrical engineering and computer science from the National Tsing Hua University, Hsinchu, Taiwan, in 2019. He is currently a Research Assistant in electrical and computer engineering with Georgia Institute of Technology, Atlanta, GA, USA. His current research interests include deep learning algorithms and algorithm-hardware co-design for deep learning.

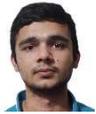
Janak Sharda received the B.Tech. degree in electrical engineering from the Indian Institute of Technology Delhi, India, in 2021. He is currently pursuing the Ph.D. degree in electrical and computer engineering with the Georgia Institute of Technology, Atlanta, GA, USA. His research interests include 2.5D/3D integration, CMOS image sensors, and designing hardware accelerators.

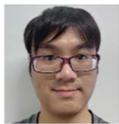
Cheng-Jhih Shih received the B.S. and M.S. degrees in computer science and information engineering from National Taiwan University. He is currently pursuing the Ph.D. degree with the School of Computer Science, Georgia Institute of Technology, Atlanta, GA, USA, under the supervision of Dr. Yingyan (Celine) Lin. His research interests include simulation, neural rendering, and hardware-software co-design.

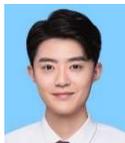
Yuyao Kong received the B.S. degree from Nanjing Tech University, Nanjing, China, in 2015, the M.S. degree from the University of Southampton, Southampton, U.K., in 2016, and the Ph.D. degree from the School of Electronic Science and Engineering, Southeast University, Nanjing, China, in 2023. He is currently a Postdoctoral Fellow with the Laboratory for Emerging Devices and Circuits, Georgia Institute of Technology, advised by Prof. Shimeng Yu. His research interests include compute-in-memory (CIM)-based algorithm-hardware co-design targeting AI processors and probabilistic computing, as well as low-voltage SRAM and other energy-efficient circuit designs.

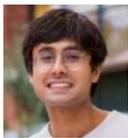
Faaiq Waqar received a B.S. degree in computer science and electrical & computer engineering from Oregon State University, Corvallis, OR, in 2022. He is currently pursuing a Ph.D. in electrical & computer engineering from the Georgia Institute of Technology, Atlanta, GA. Prior to joining Georgia Tech, he worked as a hardware engineer for Microsoft's Silicon Engineering Solutions team. He was the recipient of the NSF Graduate Research Fellowship and the Georgia Tech President's Fellowship in 2023. His current research interests pertain to the modeling and metrology of emerging amorphous oxide semiconductor and ferroelectric devices for applications in neuromorphic, reconfigurable, and high-performance computational systems.

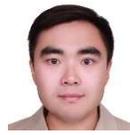
Pin-Jun Chen received the B.S. degree in Materials Science and Engineering from National Tsing Hua University, Hsinchu, Taiwan, in 2017, and the M.S. degree from the International College of Semiconductor Technology, National Chiao Tung University, Hsinchu, in 2020. He is currently pursuing the Ph.D. degree in the Department of Electrical and Computer Engineering at the Georgia Institute of Technology, Atlanta, GA, USA. His current research interests include the thermal and electrical design of 3D emerging memory-based hardware accelerators for ultra-large AI models.

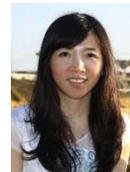
Yingyan (Celine) Lin is currently an Associate Professor in the School of Computer Science and the Co-Director of the newly established Center for Advancing Responsible Computing (CARE) at the Georgia Institute of Technology. She leads the Efficient and Intelligent Computing (EIC) Lab at Georgia Tech, which focuses on developing efficient machine learning solutions through cross-layer innovations—from efficient AI algorithms and AI hardware accelerators to AI acceleration chips—with the goal of promoting green AI and enabling ubiquitous AI-powered intelligence. She earned her Ph.D. in Electrical and Computer Engineering from the University of Illinois at Urbana-Champaign in 2017 and was an Assistant Professor at Rice University from 2017 to 2022. Celine has been recognized with multiple awards, including the Facebook Research Award, NSF CAREER Award, IBM Faculty Award, Meta Faculty Research Award (twice), ACM SIGDA Outstanding Young Faculty Award, and SRC Young Faculty Award. At Georgia Tech, she received the James D. Lester III Endowment Award in 2024 and the CoC Outstanding Mid-Career Faculty Research Award in 2025.

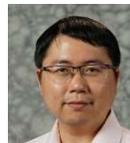
Shimeng Yu (Fellow, IEEE) is a full professor of electrical and computer engineering at Georgia Institute of Technology, where he holds the Dean's Professorship. He received the B.S. degree in microelectronics from Peking University in 2009, and the M.S. degree and Ph.D. degree in electrical engineering from Stanford University in 2011 and 2013, respectively. From 2013 to 2018, he was an assistant professor at Arizona State University. He is elevated for the IEEE Fellow for contributions to non-volatile memories and in-memory computing. His general research interests are semiconductor devices and integrated circuits for energy-efficient computing systems. His expertise is on the emerging non-volatile memories for AI hardware and 3D integration. Prof. Yu's 400+ journal/conference publications received more than 30,000 citations with H-index 82. He is the theme lead of two SRC/DARPA JUMP 2.0 centers on intelligent memory/storage and heterogeneous/monolithic 3D integration.